\documentclass[11pt,a4paper,onecolumn]{article}

\usepackage[left=2.45cm,right=2.45cm,top=2.45cm,bottom=2.45cm]{geometry}
\usepackage[T1]{fontenc}
\usepackage[utf8]{inputenc}
\usepackage{newtxtext}
\usepackage{microtype}
\usepackage{setspace}
\usepackage{authblk}
\usepackage{graphicx}
\usepackage{multirow}
\usepackage{amsmath,amsfonts,amsthm,mathtools}
\usepackage{newtxmath}
\usepackage{amsbsy}
\usepackage{dsfont}
\usepackage{braket}
\usepackage{array}
\usepackage{gensymb}
\usepackage{placeins}
\usepackage{xcolor}
\usepackage{booktabs}
\usepackage{mfirstuc}
\usepackage[version=4]{mhchem}
\usepackage{bm}
\usepackage[normalem]{ulem}
\usepackage{textcomp}
\usepackage{epstopdf}
\usepackage{xspace}
\usepackage{ragged2e}
\usepackage{titlesec}
\usepackage{indentfirst}
\usepackage[font=small,labelfont=bf,singlelinecheck=false,justification=RaggedRight]{caption}
\usepackage[super,sort&compress]{natbib}
\setcitestyle{citesep={,}}
\usepackage[colorlinks=true,linkcolor=blue,citecolor=blue,urlcolor=blue]{hyperref}

\renewcommand{\thesubsubsection}{\arabic{subsubsection}.}

\makeatletter

\renewcommand{\thesubsection}{\Alph{subsection}.}

\titleformat{\section}
  {\normalfont\large\bfseries}
  {\thesection}
  {0.6em}
  {\MakeUppercase}

\titleformat{\subsection}
  {\normalfont\normalsize\bfseries}
  {\thesubsection}
  {0.6em}
  {}

\titleformat{\subsubsection}
  {\normalfont\normalsize\itshape}
  {\thesubsubsection}
  {0.6em}
  {}
\titlespacing*{\section}{0pt}{2.2ex plus .8ex minus .2ex}{1.0ex}
\titlespacing*{\subsection}{0pt}{1.8ex plus .6ex minus .2ex}{0.7ex}
\titlespacing*{\subsubsection}{0pt}{1.4ex plus .5ex minus .2ex}{0.5ex}

\DeclareCaptionLabelSeparator{npjbar}{\enspace|\enspace}
\newcommand{\ten}[1]{\bm{#1}}

\newcommand{\vc}[1]{\boldsymbol{\mathrm #1}}

\newcommand{\AMATIS}{\texttt{AMATIS}\xspace}

\newcommand{\SpinGraph}{\texttt{SpinGraph}\xspace}

\DeclareMathOperator{\Null}{Null}

\makeatletter
\newcommand\sh@ftsym[1]{\smash{\raise-.3ex\hbox{$\scriptscriptstyle#1$}}}
\newcommand\Gtrless{\mathbin{\sh@ftsym({\gtrless}\sh@ftsym)}}
\makeatother

\makeatletter
\def\my@tag@font{\normalsize}
\def\maketag@@@#1{\hbox{\m@th\normalfont\my@tag@font#1}}
\let\amsmath@eqref\eqref
\renewcommand{\eqref}[1]{{\let\my@tag@font\relax\amsmath@eqref{#1}}}
\makeatother

\allowdisplaybreaks

\title{\bfseries Automated Generation of Commensurate Magnetic Structures based on Spin Space Groups and Graph Theory}

\author[1,2]{Yifan Wang}
\author[1,2]{Boyang Deng}
\author[3]{John Robertson}
\author[1,2]{Weisheng Zhao}
\author[4,5]{Stefan Bl\"ugel}
\author[1,2]{Haichang Lu\thanks{Correspondence: HaichangLu@buaa.edu.cn}}

\affil[1]{Fert Beijing Institute, MIIT Key Laboratory of Spintronics, School of Integrated Circuit Science and Engineering, Beihang University, Beijing, 100191, China}
\affil[2]{State Key Laboratory of Spintronics, Hangzhou International Innovation Institute, Beihang University, Hangzhou 311115, China}
\affil[3]{Engineering Department, Cambridge University, Cambridge CB2 1PZ, UK}
\affil[4]{Peter Gr\"unberg Institut, Forschungszentrum J\"ulich and JARA, 52425 J\"ulich, Germany}
\affil[5]{Institute for Theoretical Physics, RWTH Aachen University, 52074 Aachen, Germany}

\date{}

\begin{document}
\maketitle

\begin{abstract}
Magnetic structures with symmetry constraints are candidates for energetically favorable configurations. Enumerating these structures is essential for identifying experimental observations and provides unbiased, linearly stable, and optimally sampled reference configurations for energy fitting when extracting spin interactions. We present \SpinGraph, an automated workflow generating symmetry-distinct magnetic configurations. \SpinGraph is not only compatible with the more general spin space groups, but also allows precise control of the prescribed single- or multi-$Q$ states superposition. For finite groups, we enumerate compatible subgroups directly. For spin space groups with a continuous or special spin-only part, the finite component is enumerated first and then combined with the compatible exact spin-only constraints. These actions are translated into graph constraints in real or Fourier space. The symmetry of every generated structure is re-evaluated after construction. Finally, we integrate \SpinGraph with the magnetic analysis code \AMATIS to perform a thorough calculation for the spin interactions of the insulating monolayer CrI\textsubscript{3}. \SpinGraph complements the last part of \AMATIS and realizes the fully automated workflow for the spin Hamiltonian construction, which is essential for studying phase transitions, spin textures, magnons, and spin dynamics.
\end{abstract}

\section*{Introduction}

A long-standing goal of computational magnetism is to describe magnetic ground states, collective excitations, and phase transitions in agreement with experiment. Effective Hamiltonians make this problem manageable by retaining only the degrees of freedom relevant to the phenomenon of interest. For magnetism, this leads to the development of the spin model that traces back from the Heisenberg model\cite{Heisenberg1928,VanVleck1937,Anderson1950} to models with relativistic and higher-order interactions\cite{MoriyaYosida1953,Moriya1960,Kittel1960,Nussinov2015,Brinker2019}, and finally the recent general tensorial model\cite{amatis}. Once their coefficients have been obtained from first-principles calculations, these models can be transferred into various post-processing calculations. Energy mapping is a central algorithm to extract spin interaction coefficients from the Schr\"odinger electronic structure. It uses the total energies of a set of magnetic configurations to fit into the energy landscape within a predefined framework of the spin model. The quality of the fit depends heavily on the choice of the configurations. 

Symmetric magnetic structures are configurations that remain invariant under symmetry operations belonging to various groups. They are energy-degenerate points that are well-spread over the configuration space. Statistically, sampling symmetric structures can increase the linear stability of the fitting. They are also highly likely to be the local energy minimum, which is accessible via the self-consistent calculation. Even without the fitting procedure, the ground states of most magnets can be tractable via enumerating the symmetry structures and comparing their total energies. Therefore, it is of great significance to list symmetry magnetic structures, especially in an automated workflow so that density functional theory can be implemented in a high-throughput way. Previously, superpositions of spin spiral states were prepared by hand in \AMATIS\cite{amatis}, catering only to the magnetic space groups\cite{PerezMato2012MagneticSuperspace}. This way of selection already reproduces a robust fitting of the spin interactions. Here, we aim to develop an automated workflow for generating symmetric structures in the expanded framework of the spin space group, which is important for diverse magnet data construction. 

Current methods commonly involve two ways to generate symmetry magnetic structures. One imposes a magnetic space group directly and keeps the structures that obey it\cite{RodriguezCarvajalPerezMato2024MSGvsRA,Petricek2010MagneticSuperspace}, such as Bilbao Crystallographic Server and JANA2020 that construct commensurate and incommensurate structures using magnetic superspace groups\cite{Bilbao,PerezMato2015MagneticTools,PerezMato2012MagneticSuperspace,Henriques2024JANA}. The other uses representation analysis within Landau theory\cite{Bertaut1968RepresentationAnalysis,Bertaut1971MagneticGroupTheory,Bertaut1981GroupTheory,PhysRevB.76.054447}. BASIREPS implements Bertaut's representation analysis\cite{RodriguezCarvajal2025FullProf}, and SARAh provides online representation-analysis tools\cite{Wills2025SARAh}. More recently, spin-space-group symmetry has been incorporated into magnetic-structure generation. Nomoto \textit{et al.}\cite{Nomoto2026MagneticStructureSSG} constructed spin-symmetry-adapted structures from totally symmetric representations of oriented spin space groups, while Li \textit{et al.}\cite{Li2025SymmetryGuidedMagnetic} developed a symmetry-guided framework for systematically generating magnetic configurations and predicting magnetic ground states. Though these approaches provided important foundations for the recent magnet data structure constructions, there remain a few limitations for each of them. For instance, representation-analysis methods naturally incorporate prescribed propagation ($\vc{q}$) vectors, but their conventional formulations are not readily compatible with general spin-space-group symmetries. Conversely, existing spin-space-group-based methods can generate structures with prescribed spin symmetry, but do not directly provide systematic construction under specified single- or multi-$Q$ constraints. Explicit propagation-vector constraints are important because magnetic Bragg peaks may identify the relevant propagation vector(s) without uniquely determining the underlying magnetic structure\cite{RodriguezCarvajal1993Neutron,RodriguezCarvajalPerezMato2024MSGvsRA}.

We therefore developed \SpinGraph, an automated workflow that is compatible with both the aforementioned requirements of the spin space groups framework and the propagation-vector constraints. It uniquely formulates magnetic-structure generation as a graph-constrained problem, enabling an exact treatment of continuous spin-only components and a unified description of real-space, Fourier-space, supercell, and propagation-vector constraints within a single workflow. We demonstrate the method using CrI\textsubscript{3} as a representative example, where \SpinGraph systematically generates symmetry-distinct magnetic structures under different supercell and $Q$ constraints and reproduces the expected symmetry relations. Using 396 generated configurations for energy mapping, we further recover the few dominant Heisenberg exchange, anisotropic interactions, and higher-order spin interactions consistent with the established magnetic behavior of CrI\textsubscript{3}.

\section*{Results}

\subsection{SpinGraph framework}

\SpinGraph provides three schemes for calculating symmetry-constrained magnetic structures. The first constrains only the supercell size, the second constrains the propagation vectors of the magnetic structure, and the third constrains both the supercell size and the propagation vectors. The program then decomposes the continuous and discrete parts of the spin space group (SSG), combines them, and attempts to construct SSGs with different symmetry-breaking patterns. For each newly constructed SSG, the program identifies its generators and the infinitesimal generators of its Lie-group part, constructs the graph structure, and solves the linear null space. Finally, zero-total-magnetic-moment constraints or equal-magnetic-moment constraints are added to obtain the magnetic structures.

\subsubsection{Spin space group action}

An SSG is a symmetry group in which the spin and spatial parts are decoupled, and therefore has more degrees of freedom than a conventional magnetic space group (MSG)\cite{BrinkmanElliott1966SpinSpaceGroups,Chen2024SpinSpaceGroups,Jiang2024SpinSpaceGroups}. When the spin and spatial parts are locked together, the SSG reduces to an MSG. An operation $g\in G$ is written as
\begin{equation}
\begin{aligned}
&g=\eta_g\{\ten{U}_g\parallel\ten{R}_g\mid\vc{\tau}_g\},\\
&\eta_g\in\{+1,-1\},\quad\quad
\ten{O}_g=\eta_g\ten{U}_g .
\end{aligned}
\label{eq:group_action}
\end{equation}
where $\ten{R}_g$ and $\vc{\tau}_g$ act on real-space crystallographic coordinates, $\ten{U}_g\in SO(3)$ rotates the spin, and $\eta_g=-1$ marks an operation containing time reversal\cite{Jiang2024SpinSpaceGroups}. Their product $\ten{O}_g\in O(3)$ is the operation that acts on the axial magnetic moment. Let $L$ be the set of discrete vectors with integer components, with the basis of the primitive Bravais lattice, and $L_{\mathrm{sc}}\subset L$ the set of supercell translation vectors. For the finite set $\mathcal{R}_{\mathrm{sc}}=\{\vc{r}_i\}$ of magnetic sites in that supercell, the action on a site position and its magnetic moment is
\begin{equation}
\ten{R}_g\vc{r}_i+\vc{\tau}_g
=\vc{r}_j+\vc{T}_{ij}^{(g)},\quad
\vc{T}_{ij}^{(g)}\in L_{\mathrm{sc}},
\qquad
\vc{S}_j=\ten{O}_g\vc{S}_i .
\label{eq:site_and_spin_action}
\end{equation}
Here, $\vc{T}_{ij}^{(g)}\in L_{\mathrm{sc}}$ is the supercell translation relating the transformed site to representative $\vc{r}_j$, and $\vc{S}_i$ is the axial magnetic moment at $\vc{r}_i$. We use $\vc{q}\cdot\vc{T}$ for the crystallographic phase when the propagation vector $\vc{q}$ is in reciprocal-lattice units and $\vc{T}$ is a lattice translation, with the customary factor $2\pi$ included. For the complex Fourier amplitude $\vc{S}_i(\vc{q})$ at site $i$ and propagation vector $\vc{q}\in Q$,  where $Q$ is the selected set of allowed propagation vectors.
The reciprocal-space action is
\begin{equation}
\begin{aligned}
&\vc{q}'=\rho_g(\vc{q})=\ten{R}_g^{-\mathsf T}\vc{q}
\pmod{L^{*}},\\
&\vc{S}_j(\vc{q}')=
\exp\!\left[\mathrm{i}\vc{q}'\cdot\vc{T}_{ij}^{(g)}\right]
\ten{O}_g\vc{S}_i(\vc{q}).
\end{aligned}
\label{eq:fourier_action}
\end{equation}
Here, $L^{*}$ is the set of the reciprocal lattice which is one-to-one correpsonds to $L$, $\rho_g$ denotes the reciprocal-space action of $g$, and $\vc{q}'=\rho_g(\vc{q})$ is the transformed propagation vector. For a real magnetic structure, $\vc q$ and $-\vc q$ appear in pairs, and their Fourier components satisfy a complex-conjugation constraint. To treat this conjugation constraint uniformly within the graph structure, the complex Fourier components are realified into six-dimensional vectors. Supplementary Notes~1 and~4 derive the phase and compare the linear and anti-linear conventions for time reversal\cite{SupplementaryInfo,Schweizer2005TimeInversion}. 

\subsubsection{Compatible symmetry}
The three compatible groups used by the workflow are the supercell stabilizer $G_{\mathrm{sc}}$, the propagation-vector stabilizer $G_Q$, and their joint stabilizer $G_{\mathrm{sc},Q}$, defined by
\begin{equation}
\begin{aligned}
&G_{\mathrm{sc}}=\{g\in G\mid \ten{R}_gL_{\mathrm{sc}}=L_{\mathrm{sc}},\\
&\hspace{15mm}\{\ten{R}_g\mid\vc{\tau}_g\}\mathcal{R}_{\mathrm{sc}}
\equiv\mathcal{R}_{\mathrm{sc}}\pmod{L_{\mathrm{sc}}}\},\\
&G_Q=\{g\in G\mid \rho_g(Q)\equiv Q\pmod{L^{*}}\},\\
&G_{\mathrm{sc},Q}=G_{\mathrm{sc}}\cap G_Q .
\end{aligned}
\label{eq:compatible_groups}
\end{equation}
$G_{\mathrm{sc}}$ denotes the set of operations that leave the crystal structure invariant when the supercell is taken as the reference unit. Similarly, $G_Q$ contains the operations that permute the selected propagation vectors. $G_{\mathrm{sc},Q}$ denotes the set of operations allowed under both the supercell constraint and the $Q$-set constraint. In the following, $G_{\mathrm c}$ denotes the symmetry group selected in the calculation, with pure supercell translations in $L_{\mathrm{sc}}$ understood to have been quotiented out. Supplementary Note~2 gives the subgroup proofs\cite{SupplementaryInfo}.

If $G_{\mathrm{c}}^{\circ}$ is the identity component, i.e., the connected subgroup containing the identity element and all symmetry operations continuously connected to it, the discrete symmetry is represented by the finite component group,
\begin{equation}
\overline{G}_{\mathrm{c}}=G_{\mathrm{c}}/G_{\mathrm{c}}^{\circ},
\qquad
\pi:G_{\mathrm{c}}\rightarrow\overline{G}_{\mathrm{c}},
\qquad
\overline{K}=\pi(K)\leq\overline{G}_{\mathrm{c}}.
\label{eq:finite_component_group}
\end{equation}
For a finite parent group, $G_{\mathrm{c}}^{\circ}=\{e\}$ and this quotient is simply the original compatible group. We enumerate the subgroups of $\overline{G}_{\mathrm{c}}$ and then restores the compatible spin-only constraints. The subgroups of the discrete component group are enumerated using its Cayley table, and details are provided in Supplementary Note 3\cite{SupplementaryInfo}. However, a continuous parent's exact closed subgroups generally do not form a finite list\cite{Hall2015LieGroups,Sepanski2007CompactLieGroups}. The algorithm therefore separates the finite and spin-only parts. For every enumerated subgroup $\overline{K}\leq\overline{G}_{\mathrm{c}}$, it selects a spin-only subgroup $\mathcal{K}_K$ that is contained in the parent spin-only factor and is normalized by the finite subgroup,
\begin{equation}
 h\mathcal{K}_Kh^{-1}=\mathcal{K}_K,
 \qquad h\in\overline{K}.
\label{eq:continuous_kernel_compatibility}
\end{equation}
Here and below, $h\in\overline{K}$ is understood through a chosen representative in $K$. For magnetic-structure construction, spin-only subgroups that impose the same constraints on the allowed spin directions are treated as equivalent and need not be solved repeatedly. The corresponding subgroup relations are discussed in Supplementary Note~3\cite{SupplementaryInfo}.

\subsubsection{Graph constraints and solution spaces}
For a selected subgroup candidate $K\leq G_{\mathrm{c}}$, let $\Gamma_K^{\mathrm{disc}}=\{g_1,\ldots,g_{n_K}\}$ denotes the generators of its finite representative. We write $\vc{S}$ for a magnetic configuration and $\mathcal{D}(g)$ for the linear action of $g$. If
\begin{equation}
\mathcal{D}(g_i)\vc{S}=\vc{S},
\qquad
\forall g_i\in \Gamma_K^{\mathrm{disc}},
\label{eq:generator_invariance}
\end{equation}
then the same relation holds for every product of these generators and their inverses. In particular,
\begin{equation}
\begin{aligned}
g_{\mathrm d}=g_{i_1}^{\epsilon_1}g_{i_2}^{\epsilon_2}\cdots g_{i_{N_g}}^{\epsilon_{N_g}},
\qquad \epsilon_{\kappa}=\pm1, \\
\mathcal{D}(g_{\mathrm d})\vc{S}
=\mathcal{D}(g_{i_1}^{\epsilon_1})\cdots
\mathcal{D}(g_{i_{N_g}}^{\epsilon_{N_g}})\vc{S}
=\vc{S}.
\end{aligned}
\end{equation}
If $K$ has a nontrivial identity component $K^{\circ}$, its connected spin-only action is imposed through the Lie algebra
\begin{equation}
\begin{aligned}
&\mathfrak{k}=\operatorname{Lie}(K^{\circ})
=\operatorname{span}_{\mathbb{R}}
\{\ten{X}_1,\ldots,\ten{X}_{r_K}\},\\
&\mathcal{L}_a\vc{S}=\vc{0}
\quad(a=1,\ldots,r_K),
\end{aligned}
\label{eq:continuous_generator_constraint}
\end{equation}
where $\mathcal{L}_a$ is the infinitesimal action induced by $\ten{X}_a$. $SO(2)$ requires one independent generator, whereas $SO(3)$ requires three\cite{LitvinOpechowski1974SpinGroups,Liu2022SpinGroupSymmetry,Schiff2025SpinPointGroups}. Coplanar constraints are imposed directly by projecting the spins onto the selected plane.

The discrete generator actions define either a real-space graph $\mathcal{G}_{\mathrm r}$ or a Fourier-space graph $\mathcal{G}_{\mathrm q}$. Let $\vc{x}_i$ be the variable at vertex $i$. Each directed edge $e$ stores a matrix $\ten{P}_e$ relating its source $s(e)$ to its target $t(e)$,
\begin{equation}
\vc{x}_{t(e)}=\ten{P}_e\vc{x}_{s(e)}.
\label{eq:edge_constraint}
\end{equation}
In real space, $\vc{x}_i=\vc{S}_i\in\mathbb{R}^3$. In Fourier space, it is the $\mathbb{R}^6$ form of a complex amplitude. In $\mathcal{G}_{\mathrm r}$, each site $\vc{r}_i\in\mathcal{R}_{\mathrm{sc}}$ is a vertex, and Eq.~\eqref{eq:site_and_spin_action} supplies the edge matrix $\ten{O}_g$. In $\mathcal{G}_{\mathrm q}$, a vertex is a pair $(i,\vc{q})$, and the edge matrix also contains the phase in Eq.~\eqref{eq:fourier_action}. Using the constructed graph and the transport relations, linear constraints can be obtained by identifying the fundamental cycles of the graph. A graph may contain multiple connected components, and independent degrees of freedom are not shared between different components, whereas vertices within the same component share the same set of independent degrees of freedom. For each connected component, a spanning tree transports a reference variable $\vc{x}_0$ through the component, $\vc{x}_i=\ten{M}_i\vc{x}_0$, while each remaining edge $e:i\rightarrow j$ imposes a cycle-consistency condition\cite{Diestel2017GraphTheory,Cormen2022Algorithms}. Stacking these conditions together with the continuous and planar constraints gives
\begin{equation}
\bigl(\ten{M}_j-\ten{P}_e\ten{M}_i\bigr)\vc{x}_0=\vc{0},
\qquad
\ten{N}_{\mathrm{comp}}=\Null(\ten{A}_{\mathrm{comp}}),
\qquad
\ten{B}_i=\ten{M}_i\ten{N}_{\mathrm{comp}}.
\label{eq:component_basis}
\end{equation}
The third equation shows that the basis vectors at the other vertices within a connected component can be obtained through the transport relations and are not independent. Supplementary Note~5 gives the group constraints and fundamental-cycle details\cite{Paton1969FundamentalCycles,Deo1982FundamentalCycles,SupplementaryInfo}.

\subsubsection{Fourier-space restriction and projection}
Let $V_{\mathrm{sc}}$ be the space of all magnetic configurations in the selected supercell, and let $V_Q\subseteq V_{\mathrm{sc}}$ contain only the Fourier components in $Q$. In a real-space calculation restricted to $Q$, the projector $P_Q:V_{\mathrm{sc}}\rightarrow V_Q$ removes all other Fourier components. Because both the supercell constraint and the $Q$-set constraint must be satisfied, the corresponding symmetry group is $G_{\mathrm{sc},Q}$, and the resulting solution space is
\begin{equation}
P_Q V_{\mathrm{sc}}^{G_{\mathrm{sc},Q}}=V_Q^{G_{\mathrm{sc},Q}}.
\label{eq:q_restricted_solution}
\end{equation}
Thus, projecting the real-space solution onto $Q$ gives the same solution space as solving directly in Fourier space. The equivalence is proved in Supplementary Notes~6 and~7\cite{SupplementaryInfo}.

Although the real-space calculation is performed using $G_{\mathrm{sc},Q}$, the resulting magnetic structures may still contain propagation vectors outside $Q$. Therefore, the basis obtained from the real-space calculation must be projected onto the allowed subspace using the projection operator. Thus, the real-space basis is projected as
\begin{equation}
\widetilde{\ten{B}}_{\mathrm{sc},Q}=P_Q\ten{B}_{\mathrm{sc},Q},
\qquad
\vc{S}_Q=\widetilde{\ten{B}}_{\mathrm{sc},Q}\vc{c}.
\label{eq:q_projected_basis_constraint}
\end{equation}

\subsubsection{Construction of physical magnetic structures}

After obtaining the symmetry-allowed basis, additional physical constraints are imposed directly on its global coefficients. A zero-total-spin condition further reduces the allowed coefficient space, and an equal-moment-length condition constrains the remaining coefficients. In contrast, normalizing each site independently may break these relations or introduce Fourier components outside $Q$. The detailed procedure is given in Supplementary Note~8\cite{SupplementaryInfo}.

The symmetry used to construct a magnetic structure is not necessarily its full symmetry, because a particular solution may have additional operations. Therefore, after a structure is generated, we should check all allowed spatial and spin operations and determine its maximal stabilizer $K_{\max}$, after which the full symmetry operations are verified against the magnetic structure\cite{Chen2024SpinSpaceGroups,Xiao2024SpinSpaceGroups,Liu2026OrientedSpinSpaceGroups,yu2026identifyingorientedspinspace}.

\subsection{SpinGraph workflow}

\begin{figure}[t]
    \centering
    \includegraphics[width=\linewidth]{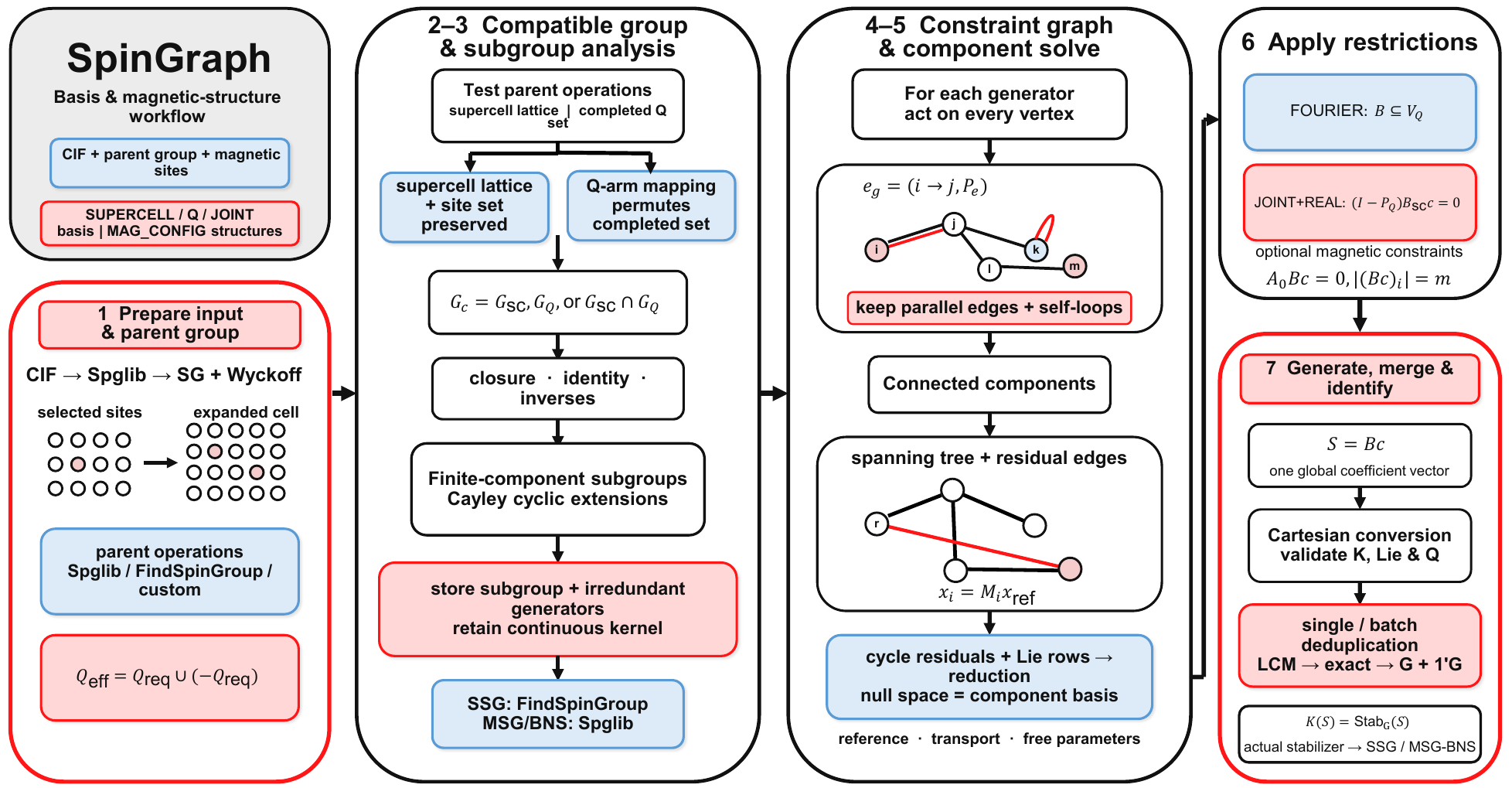}
    \caption{\textbf{Overall workflow of \SpinGraph}. The crystal structure, selected magnetic sites, supercell, and prescribed propagation-vector set $Q$ define a finite magnetic problem. \SpinGraph finds the compatible group and its subgroups, converts their actions into graph constraints, and solves each connected component. It then restricts the basis to the requested Fourier components, reconstructs the magnetic structures when needed, and checks them against the full symmetry. The program uses \texttt{spglib} and \texttt{findspingroup} for crystallographic symmetry analysis\cite{Togo2024Spglib,yu2026identifyingorientedspinspace}.}
    \label{fig:workflow}
\end{figure}

The inputs are a crystal structure, the magnetic sites, the spin space group operations, and a supercell, a propagation-vector set $Q$, or both. The output contains the allowed magnetic basis together with the subgroup and graph information used to obtain it. Figure~\ref{fig:workflow} summarizes the following steps.

\begin{enumerate}

\item \textit{Prepare the input data.}
\SpinGraph reads the CIF, determines the crystallographic symmetry and Wyckoff positions with Spglib, and expands the selected magnetic sites into a supercell\cite{Togo2024Spglib}. The parent magnetic or spin space group operations are then obtained from Spglib, FindSpinGroup, or a user-supplied SSG file\cite{yu2026identifyingorientedspinspace}. For Fourier calculations, the program also completes the input propagation-vector set for real conditions.

\item \textit{Select the compatible operations.}
Each parent operation is tested using Eq.~\eqref{eq:compatible_groups}. Depending on the calculation mode, it gives $G_{\mathrm{sc}}$, $G_Q$, or $G_{\mathrm{sc},Q}$. The program verifies the identity, closure, and inverses of the group. For a continuous SSG, the finite component is used here, while the continuous spin constraints are retained separately.

\item \textit{Enumerate the subgroups.}
The finite subgroups are enumerated by cyclic extension of the Cayley table\cite{Hulpke2018Subgroups,HoltEickOBrien2005ComputationalGroups,Cannon2001Subgroups}. For a continuous parent, this step is applied to its finite component group, and each finite subgroup is then combined with the spin-only axial, plane-reflection, or trivial constraint allowed by Eq.~\eqref{eq:continuous_kernel_compatibility}. Then the corresponding groups are identified by Spglib or FindSpinGroup.

\item \textit{Build the constraint graph.}
The generators in $\Gamma_K^{\mathrm{disc}}$ are applied to all graph vertices. Real-space edges relate magnetic moments on different sites, while Fourier-space edges relate amplitudes at different sites and propagation vectors. Continuous spin symmetries are added as exact local linear constraints as self-loops.

\item \textit{Solve the graph constraints.}
The graph is divided into connected components\cite{Diestel2017GraphTheory,Cormen2022Algorithms}. A spanning tree, generated by BFS\cite{Cormen2022Algorithms}, transports a reference vector through each component, and the remaining edges provide a complete set of consistency conditions. Solving these conditions gives the allowed real-space or Fourier-space basis and its free parameters.

\item \textit{Apply the additional restrictions.}
For a joint calculation $G_{\mathrm{sc},Q}$ in the real-space representation, \SpinGraph projects the solved basis onto the requested propagation vectors and removes zero or dependent basis vectors. When magnetic structures are generated, optional zero-total-spin and equal-moment constraints are imposed on the complete global coefficient space. Only one common scaling factor is applied, while sitewise normalization is not used.

\item \textit{Generate and validate the output.}
The component bases are combined into the full magnetic basis, and generated moments are transformed to Cartesian coordinates. \SpinGraph checks the complete subgroup, the continuous constraints, the prescribed Fourier support, and any enabled magnetic constraints. After equivalent structures are removed, the stabilizer of each magnetic configuration is recomputed.

\end{enumerate}

\subsection{Magnetic-structure generation and validation}
\subsubsection{Magnetic-structure generation using a \texorpdfstring{CrI\textsubscript{3}}{CrI3} monolayer}

We compare the three calculation modes using the same monolayer CrI\textsubscript{3} structure. It belongs to $P\bar{3}1m$ (No.~162), and the two Cr atoms at the $2d$ Wyckoff position, $\vc{r}_1=(1/3,2/3,1/2)$ and $\vc{r}_2=(2/3,1/3,1/2)$, are chosen as magnetic sites. The parent symmetry is spin space group 162.157.8.1. For a direct comparison, we use the same embedded spin space group 10.6.1.3 in all three cases.

In $G_{\mathrm{sc}}$ , we use a $(3\times1\times1)$ supercell and place no restriction on $Q$, and the Fourier decomposition may contain $\vc{q}=(0,0,0)$, $(1/3,0,0)$, and $(-1/3,0,0)$. For spin space group 10.6.1.3, the six Cr sites form three independent pairs. Each pair has an in-plane basis direction $\vc{u}_{\mathrm{sc}}=(\kappa,1,0)$, where $\kappa=(1+\sqrt{3}+\sqrt{6})/2$, and an out-of-plane direction $\vc{e}_z=(0,0,1)$. This gives six real free parameters, as listed in Table~\ref{tab:cri3_supercell}.

\begin{table}[htbp]
    \centering
    \caption{\textbf{$G_{\mathrm{sc}}$ result for parent spin space group 162.157.8.1.} The table gives the real-space basis of spin space group 10.6.1.3. Here $\vc{u}_{\mathrm{sc}}=((1+\sqrt{3}+\sqrt{6})/2,1,0)$, $\vc{e}_z=(0,0,1)$, and $a_0,a_1,a_2,b_0,b_1,b_2$ are independent real coefficients. The label $\mathrm{Cr}_{\mu\ell}$ denotes primitive-cell site $\mu$ in translation copy $\ell$.
}

    \label{tab:cri3_supercell}
    \footnotesize
    \setlength{\tabcolsep}{8pt}
    \begin{tabular}{@{}c c c@{}}
        \toprule
        Site
        & Coordinate
        & Moment $\vc{S}_{\mu\ell}$ \\
        \midrule
        $\mathrm{Cr}_{10}$
        & $\left(1/3,2/3,1/2\right)$
        & $a_0\vc{u}_{\mathrm{sc}}+b_0\vc{e}_z$ \\

        $\mathrm{Cr}_{11}$
        & $\left(4/3,2/3,1/2\right)$
        & $a_1\vc{u}_{\mathrm{sc}}+b_1\vc{e}_z$ \\

        $\mathrm{Cr}_{12}$
        & $\left(7/3,2/3,1/2\right)$
        & $a_2\vc{u}_{\mathrm{sc}}+b_2\vc{e}_z$ \\

        $\mathrm{Cr}_{20}$
        & $\left(2/3,1/3,1/2\right)$
        & $a_2\vc{u}_{\mathrm{sc}}+b_2\vc{e}_z$ \\

        $\mathrm{Cr}_{21}$
        & $\left(5/3,1/3,1/2\right)$
        & $a_1\vc{u}_{\mathrm{sc}}+b_1\vc{e}_z$ \\

        $\mathrm{Cr}_{22}$
        & $\left(8/3,1/3,1/2\right)$
        & $a_0\vc{u}_{\mathrm{sc}}+b_0\vc{e}_z$ \\
        \bottomrule
    \end{tabular}
    
\end{table}

In $G_{Q}$, we choose $\vc{q}=(1/3,0,0)$, so the effective set is $Q=\{\vc{q},-\vc{q}\}$. Spin space group 10.6.1.3 has one connected component and four real free parameters. Here, we define $\vc{A}=\alpha\vc{u}_{\mathrm{sc}}+\beta\vc{e}_z$, $\vc{B}=\gamma\vc{u}_{\mathrm{sc}}+\delta\vc{e}_z$, and $\omega=e^{2\pi\mathrm{i}/3}$. The amplitudes at $\vc{q}$ are given in Table~\ref{tab:cri3_fourier}, while $\vc{S}_{\mu}(-\vc{q})=\vc{S}_{\mu}(\vc{q})^{*}$. The real-space spin is reconstructed as $\vc{S}_{\mu\ell}=\sum_{\vc{q}\in Q}\vc{S}_{\mu}(\vc{q})e^{-\mathrm{i}\vc{q}\cdot\vc{T}_{\ell}}$, where $\vc{T}_{\ell}$ is the translation of cell $\ell$. The conjugate pair makes the reconstructed spin real and removes the $\vc{q}=(0,0,0)$ component.

\begin{table}[htbp]
    \centering
    \caption{\textbf{$G_Q$ result for parent spin space group 162.157.8.1.} The table gives the Fourier basis of spin space group 10.6.1.3 at $\vc{q}=(1/3,0,0)$. Here $\vc{A}=\alpha\vc{u}_{\mathrm{sc}}+\beta\vc{e}_z$, $\vc{B}=\gamma\vc{u}_{\mathrm{sc}}+\delta\vc{e}_z$, $\vc{u}_{\mathrm{sc}}=((1+\sqrt{3}+\sqrt{6})/2,1,0)$, and $\omega=e^{2\pi\mathrm{i}/3}$. The four coefficients $\alpha,\beta,\gamma,\delta$ are independent and real.
}
    \label{tab:cri3_fourier}
    \footnotesize
    \setlength{\tabcolsep}{8pt}
    \begin{tabular}{@{}c c c@{}}
        \toprule
        Site & Coordinate & $\vc{S}_\mu(\vc{q})$  \\
        \midrule
        $\mathrm{Cr}_1$ & $(1/3,2/3,1/2)$
        & $\vc{A}+\mathrm{i}\vc{B}$
         \\
        $\mathrm{Cr}_2$ & $(2/3,1/3,1/2)$
        & $\omega^2\left(\vc{A}-\mathrm{i}\vc{B}\right)$
         \\
        \bottomrule
    \end{tabular}
\end{table}

In $G_{\mathrm{sc},Q}$, we use the same $(3\times1\times1)$ supercell, $Q=\{\vc{q},-\vc{q}\}$. The three Cr-site pairs found in $G_{\mathrm{sc}}$ remain, but the projection couples their coefficients and reduces the basis from six to four real parameters. Equivalently, the three moments on each Cr sublattice sum to zero, so the $\vc{q}=(0,0,0)$ component vanishes. Table~\ref{tab:cri3_joint} lists the resulting real-space basis. The $G_{Q}$ calculation imposes the restriction directly in Fourier space, while the $G_{\mathrm{sc},Q}$ calculation obtains the same restricted space by projecting the real-space basis, since $G_{Q}=G_{\mathrm{sc},Q}$ in this case.

\begin{table}[htbp]
    \centering
    \caption{\textbf{$G_{\mathrm{sc},Q}$ result for parent spin space group 162.157.8.1.} The table gives the real-space basis of spin space group 10.6.1.3 for a $3\times1\times1$ supercell and $Q=\{(1/3,0,0),(-1/3,0,0)\}$. Here $\vc{u}_{\mathrm{sc}}=((1+\sqrt{3}+\sqrt{6})/2,1,0)$, $\vc{e}_z=(0,0,1)$, and $a,b,c,d$ are independent real coefficients.}
    \label{tab:cri3_joint}
    \footnotesize
    \setlength{\tabcolsep}{7pt}
    \begin{tabular}{@{}c c c@{}}
        \toprule
        Site & Coordinate & Moment $\vc{S}_{\mu\ell}$ \\
        \midrule
        $\mathrm{Cr}_{10}$ & $(1/3,2/3,1/2)$ & $-(a+c)\vc{u}_{\mathrm{sc}}-(b+d)\vc{e}_z$ \\
        $\mathrm{Cr}_{11}$ & $(4/3,2/3,1/2)$ & $a\vc{u}_{\mathrm{sc}}+b\vc{e}_z$ \\
        $\mathrm{Cr}_{12}$ & $(7/3,2/3,1/2)$ & $c\vc{u}_{\mathrm{sc}}+d\vc{e}_z$ \\
        $\mathrm{Cr}_{20}$ & $(2/3,1/3,1/2)$ & $c\vc{u}_{\mathrm{sc}}+d\vc{e}_z$ \\
        $\mathrm{Cr}_{21}$ & $(5/3,1/3,1/2)$ & $a\vc{u}_{\mathrm{sc}}+b\vc{e}_z$ \\
        $\mathrm{Cr}_{22}$ & $(8/3,1/3,1/2)$ & $-(a+c)\vc{u}_{\mathrm{sc}}-(b+d)\vc{e}_z$ \\
        \bottomrule
    \end{tabular}
\end{table}

\FloatBarrier

\subsubsection{Multi-Wyckoff real-space validation using \texorpdfstring{Fe\textsubscript{4}GeTe\textsubscript{2}}{Fe4GeTe2}}

This example checks whether the algorithm keeps inequivalent magnetic Wyckoff orbits separate in one calculation. The Fe sublattice of Fe\textsubscript{4}GeTe\textsubscript{2} in a $(1\times1\times1)$ cell has two orbits, Fe1 and Fe2, each containing six sites.

The chosen group $G_{\mathrm{c}}$ has BNS label 166.101. Its real-space graph $\mathcal{G}_{\mathrm r}$ contains all 12 Fe sites but splits into two six-site components, $\mathcal{C}_1$ and $\mathcal{C}_2$, for Fe1 and Fe2. Symmetry relates the sites within each component, but no graph edge relates the amplitudes of the two inequivalent orbits. All Fe moments within one component are therefore parallel to the crystallographic $c$ axis, while $a_1$ and $a_2$ remain independent. Table~\ref{tab:fe4gete2_structure} shows the result site by site.

\begin{table}[htbp]
    \centering
    \caption{\textbf{Orbit-resolved Fe moments in the multi-Wyckoff test.} The two inequivalent Fe orbits are listed with their symmetry-related coordinates and the moment relations for BNS 166.101. Their independent amplitudes are $a_1$ and $a_2$. The table tests whether the program keeps the two orbits separate and should not be read as a predicted magnetic ground state.}

    \label{tab:fe4gete2_structure}
    \footnotesize
    \setlength{\tabcolsep}{8pt}
    \begin{tabular}{c c c}
        \toprule
        Site & Fractional coordinate & Magnetic moment \\
        \midrule
        $\mathrm{Fe1}_1$ & $\left(0,0,0.88813\right)$
        & $\left(0,0,a_1\right)$ \\
        $\mathrm{Fe1}_2$ & $\left(0,0,0.11187\right)$
        & $\left(0,0,a_1\right)$ \\
        $\mathrm{Fe1}_3$ & $\left(2/3,1/3,0.22147\right)$
        & $\left(0,0,a_1\right)$ \\
        $\mathrm{Fe1}_4$ & $\left(2/3,1/3,0.44520\right)$
        & $\left(0,0,a_1\right)$ \\
        $\mathrm{Fe1}_5$ & $\left(1/3,2/3,0.55480\right)$
        & $\left(0,0,a_1\right)$ \\
        $\mathrm{Fe1}_6$ & $\left(1/3,2/3,0.77853\right)$
        & $\left(0,0,a_1\right)$ \\
        \midrule
        $\mathrm{Fe2}_1$ & $\left(0,0,0.81227\right)$
        & $\left(0,0,a_2\right)$ \\
        $\mathrm{Fe2}_2$ & $\left(0,0,0.18773\right)$
        & $\left(0,0,a_2\right)$ \\
        $\mathrm{Fe2}_3$ & $\left(2/3,1/3,0.14561\right)$
        & $\left(0,0,a_2\right)$ \\
        $\mathrm{Fe2}_4$ & $\left(2/3,1/3,0.52106\right)$
        & $\left(0,0,a_2\right)$ \\
        $\mathrm{Fe2}_5$ & $\left(1/3,2/3,0.47894\right)$
        & $\left(0,0,a_2\right)$ \\
        $\mathrm{Fe2}_6$ & $\left(1/3,2/3,0.85439\right)$
        & $\left(0,0,a_2\right)$ \\
        \bottomrule
    \end{tabular}
    
\end{table}

However, Table~\ref{tab:fe4gete2_structure} is not a prediction of the magnetic ground state. Symmetry fixes the allowed directions but not the values or relative sign of $a_1$ and $a_2$, and it does not compare the energies of the allowed states. That would require an energy calculation or experimental input. Here the result only shows that the program treats the two Fe orbits separately and combines their allowed moments correctly.

\subsection{Spin interactions in monolayer CrI\textsubscript{3}}

\begin{figure}[t]
    \centering
    \includegraphics[width=\linewidth]{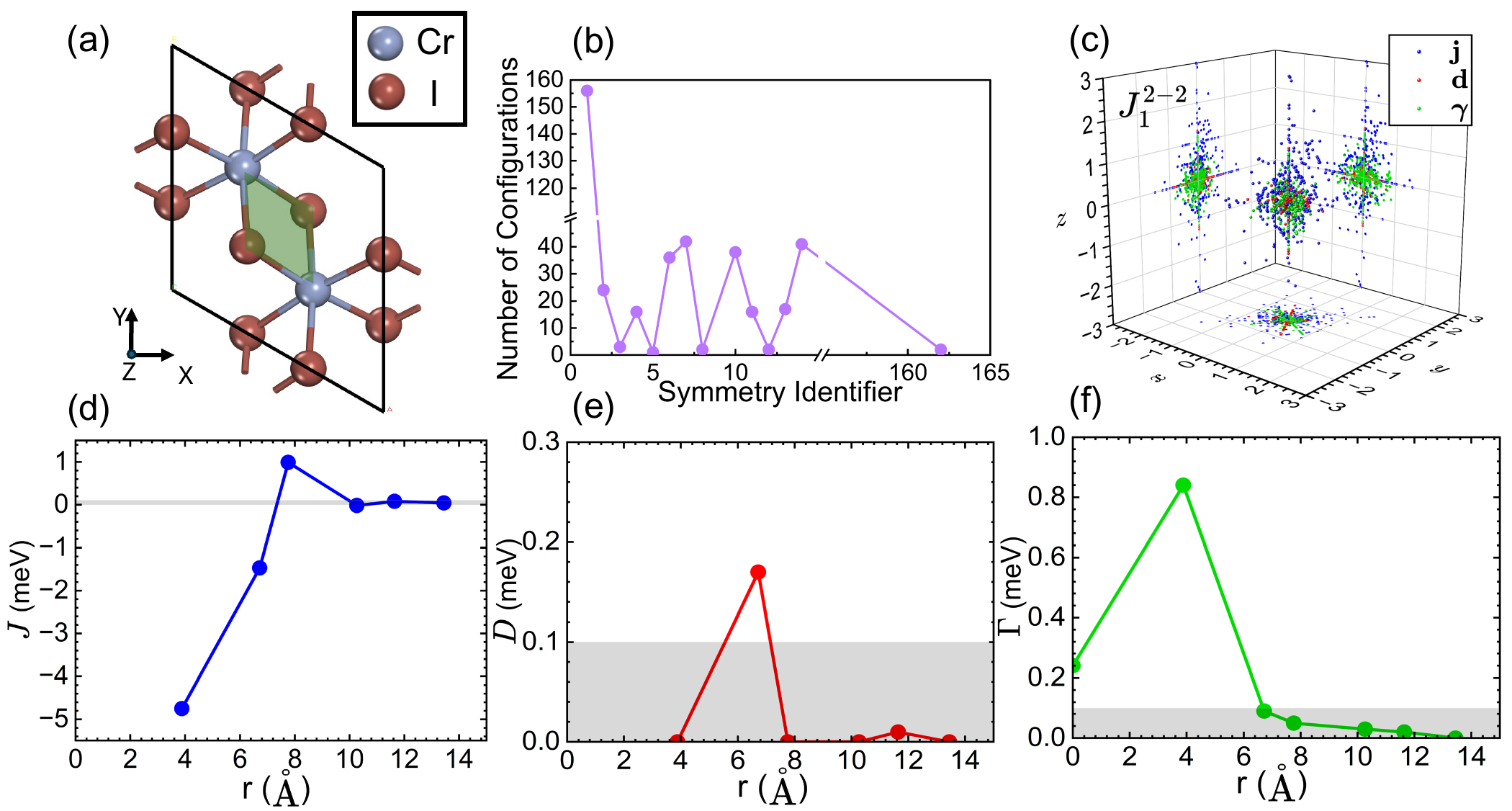}

    \caption{\textbf{Spin interactions in monolayer CrI\textsubscript{3}.} 
    \textbf{a,} Schematic illustration of the primitive-cell structure of CrI\textsubscript{3}, which crystallizes in space group No. 162, with Cr atoms occupying the magnetic Wyckoff position $2d$. The green plaquette indicates the local Cr-I-Cr-I rhombus plane associated with the nearest-neighbor Kitaev-like interaction. \textbf{b,} Number of magnetic configurations generated for different symmetries, classified by their Belov–Neronova–Smirnova (BNS) identifiers or SSG identifiers. The numerical order of these identifiers does not indicate a symmetry hierarchy. The detailed distribution is provided in the Supplementary Table~1\cite{SupplementaryInfo}. \textbf{c,} Projections of the nearest-neighbor two-ion rank-2 tensor $J_1^{2-2}$ onto the configuration subspaces. More generally, $J_p^{n-m}$ denotes the $p$th rank-$n$ interaction tensor involving $m$ sites. For multisite tensors, the index $p$ is assigned in ascending order of intersite distance. The vectors $\vc j$, $\vc d$, and $\vc \gamma$ collect the diagonal, antisymmetric off-diagonal, and symmetric off-diagonal configuration coefficients, respectively. Their component definitions and the full tensor-naming convention are provided in Supplementary Note 1\cite{SupplementaryInfo}. \textbf{d--f,} $J$ denotes the isotropic Heisenberg exchange, $D$ the Euclidean norm of the Dzyaloshinskii–Moriya vector, and $\Gamma$ the Frobenius norm of the traceless symmetric exchange matrix $\ten{\Gamma}^{ij}$ normalized by $1/\sqrt{3}$. The definitions of these scalar quantities and the tensor-naming convention are provided in Supplementary Note 1\cite{SupplementaryInfo}. The gray-shaded regions indicate the fitting error of $0.10$ meV per unit cell.
}
    \label{fig:CrI3_space}
\end{figure}

The magnetic configurations generated by \SpinGraph are used as input to \AMATIS to extract the tensorial spin interactions. We consider monolayer CrI\textsubscript{3}, a two-dimensional honeycomb insulator with space group $P\bar{3}1m$ (No.~162). The magnetic Cr atoms occupy the $2d$ Wyckoff position, with spin $S=\frac{3}{2}$.

For the rank-2 interactions, the cutoff radius is set to $r_{\mathrm c}=13~\text{\AA}$. This includes the single-ion anisotropy tensor $J_{1}^{2-1}$ and five two-site tensors up to $J_{5}^{2-2}$. For the rank-4 interactions, we include two-site tensors in which each site appears twice, up to the second-nearest-neighbor tensor $J_{2}^{4-2}$. No rank-4 single-ion term is included because $S<2$. The spin model is fitted using the 396 magnetic configurations generated above. The smallest supercell commensurate with all these configurations is $4\times2\times1$, and the total energies of all configurations are therefore calculated using this supercell. The fitted rank-2 and rank-4 tensor components are provided in the Supplementary Note 10\cite{SupplementaryInfo}, which is consistent with ref\cite{amatis}.

The extracted interactions are consistent with the experimental description of monolayer CrI\textsubscript{3} as an Ising-type ferromagnet\cite{Huang2017}. Both the single-ion anisotropy, $0.52$~meV per Cr, and the two-ion anisotropy, $0.43$~meV per Cr, favor out-of-plane spin polarization. The first three nearest-neighbor Heisenberg interactions are dominant and give a Curie temperature of about $45$~K. In addition to the two-ion anisotropy, the symmetric exchange $\ten{\Gamma}^{ij}$ contains a bond-dependent Kitaev-like interaction. For the nearest-neighbor bond, its preferred plane is the Cr-I-Cr-I rhombus shown by the green plaquette in Fig.~\ref{fig:CrI3_space} (a). Although inversion symmetry forbids a net Dzyaloshinskii--Moriya interaction, local inversion-symmetry breaking along second-nearest-neighbor Cr bonds allows antisymmetric exchange interactions of about $0.20$~meV.

\section*{Discussion}

\SpinGraph provides a comprehensive workflow for generating commensurate magnetic structures, capable of producing a series of magnetic structures with different symmetry-breaking patterns. This method can simultaneously handle both spin space groups and magnetic space groups, and can properly treat the continuous components of these groups. Furthermore, the algorithm transforms symmetry constraints into graph constraints and constructs the corresponding linear null-space problem for solution, thereby avoiding the need to establish systems of linear equations containing a large number of redundant relations, while also offering good interpretability.

During the generation of magnetic structures, constraints such as zero total magnetic moment and equal moment magnitude can be further introduced to ensure that the resulting structures meet the corresponding computational requirements. The algorithm's different computational modes share the same underlying graph structure: the real-space mode handles supercell periodicity, the Fourier mode handles a specified propagation-vector set $Q$, and the combined mode applies both types of constraints simultaneously. Consequently, this method can generate magnetic structures that simultaneously satisfy strict propagation-vector constraints and symmetry requirements, thereby covering cases that are difficult for traditional magnetic-structure generation programs to handle directly.

The final result is a space of candidate magnetic structures permitted under the given symmetry conditions, rather than the energetic ground state. The relative stability of the candidate structures still requires further evaluation through spin models, first-principles calculations, or experiments. In the future, spin superspace groups could be introduced to extend this framework to the generation of incommensurate magnetic structures.

\section*{Methods}
\setcounter{subsection}{0}
\subsection{Symmetry identification and databases}

Crystal symmetries and Wyckoff positions are identified using Spglib\cite{Togo2024Spglib}. When the spin-space symmetry reduces to an ordinary magnetic space group, Spglib is also used to identify the corresponding magnetic subgroup and the magnetic symmetry of the final generated structure. For spin-space group calculations, the parent SSG is constructed from the spin-space group database implemented in FindSpinGroup\cite{yu2026identifyingorientedspinspace}. The same database and identification routines are then used to identify the generated subgroups and the final spin-space group symmetry of each magnetic structure.

\subsection{First-principles calculations}

The \emph{ab initio} calculations were performed using the Vienna \emph{ab initio} Simulation Package (VASP)\cite{VASP1,VASP2} with the projector augmented-wave (PAW) method\cite{PAW1,PAW2}. The Cr $3p$, $3d$, and $4s$ orbitals and the I $5s$ and $5p$ orbitals were treated as valence states. A plane-wave cutoff energy of $520$~eV and a $2\times4\times1$ $k$-point mesh were used for monolayer CrI\textsubscript{3}. The electronic energy convergence criterion was set to $10^{-7}$~eV, and spin--orbit coupling and noncollinear magnetism were included.

The exchange-correlation functional was treated within the local spin-density approximation (LSDA)\cite{Vosko1980}, together with a Hubbard $U=0.5$~eV applied to the Cr $d$ orbitals, following Refs.~\cite{Xu2018,Xu2020a}. The experimental in-plane lattice constant $a=6.72$~\AA\ was used\cite{Huang2017,Xu2020a}. The out-of-plane lattice parameter was set to $c=30$~\AA, providing a vacuum spacing exceeding $20$~\AA\ to avoid interactions between periodic layers. Atomic positions were relaxed until the residual forces were below $0.03$~eV/\AA.

A $4\times2\times1$ supercell was used for the total-energy calculations. The noncollinear magnetic configurations generated by \SpinGraph were imposed using constrained magnetic moments\cite{Ma2015}, and the resulting total energies were used for the subsequent spin-interaction fitting.

\subsection{\AMATIS calculations for monolayer CrI\textsubscript{3}}
We extract the spin interactions from the generalized spin Hamiltonian using AMATIS (Automated Magnetism Analysis via Tensorial Interacting Spins)\cite{amatis,amatis_code}. The symmetry constraints of $P\bar{3}1m$ (No. 162) and the spin power limit are switched on. The truncated length of the rank-2 tensorial interaction, including spin-orbit coupling (SOC), is set to 12\AA, up to $J_5^{2-2}$; we also consider the non-relativistic part of $J_6^{2-2}$. This setting ensures that as many interactions as possible are taken into account for a supercell of $4\times2\times1$. For rank-4 tensors, the two-ion interactions with both site multiplicities of 2 are taken into account, up to the second-nearest-neighbor $J_2^{4-2}$. We use 396 \SpinGraph-generated symmetric magnetic configurations to calculate the total energies from first principles. The choice of the truncated lengths as well as the number of configurations for the energy fitting is already validated by the two-sided convergence tests\cite{amatis}, yielding the genuine systematic error of 0.10meV per unit cell. 

\section*{Data availability}

The data generated and analysed during this study are available from the corresponding author upon reasonable request.

\section*{Code availability}
The source code, manual, and tutorial package of \SpinGraph are released in Zenodo ( \href{https://doi.org/10.5281/zenodo.21947648}{10.5281/zenodo.21947648}). The archived code will be made available to the editors and reviewers during peer review upon request. A public GitHub repository will be released after submission, and the Zenodo record will serve as the persistent archived version of the code associated with this work.

\section*{Acknowledgements}
H.L. acknowledges financial support from the National Key R\&D Programs of China (No. 2025YFA1411000 and 2024YFA1410200), the Beijing Natural Science Foundation (No. 2232055), and the National Natural Science Foundation of China (No. 12204027). S.B. acknowledges financial support from the European Research Council (ERC) grant 856538 (project "3D MAGIC") and from the Deutsche Forschungsgemeinschaft (DFG, German Research Foundation) through SFB 1238 (Project C1).

\section*{Author contributions}

H.L. conceived the idea and supervised the project. Y.W. developed the theoretical framework, implemented the \SpinGraph code, analysed the results, and wrote the manuscript. B.D. performed the first-principles calculations and contributed the corresponding computational data. S.B. reviewed the manuscript and provided critical feedback. J.R. and W.Z. provided scientific guidance and contributed to discussions. All authors discussed the results, reviewed the manuscript, and approved the submitted version.

\section*{Competing interests}

The authors declare no competing interests.

\bibliographystyle{naturemag}
\bibliography{myreference}

\end{document}